\documentclass{fundam}

\usepackage{url}
\usepackage[ruled,lined]{algorithm2e}
\usepackage{graphicx}
\usepackage{tikz}
\usetikzlibrary{automata, positioning, arrows}
\usepackage{graphicx}
\usepackage[curve,frame,line,arrow,matrix]{xy}
\usepackage{tabularx}
\usepackage{rotating}
\usepackage[section]{placeins}
\usepackage{longtable}
\usepackage{pythontex}
\usepackage{listings}
\usepackage[utf8]{inputenc}
\usepackage{fancyvrb}

\usepackage{pgfplots}
\usepackage{pgfplotstable}
\pgfplotsset{compat=1.17}

\begin{document}

\title{The Rough Topology for Numerical Data}

\address{Department of Mathematics, Istanbul Medeniyet University, H1-20, Istanbul, TURKEY 34700}

\author{U\u{G}UR Y\.{I}\u{G}\.{I}T\\
Department of Mathematics \\
Istanbul Medeniyet University\\
ugur.yigit{@}medeniyet.edu.tr} 

\maketitle

\runninghead{Yiğit, U.}{The Rough Topology for Numerical Data}

\begin{abstract}
  In this paper, we generalize the rough topology and the core to numerical data by classifying objects in terms of the attribute values. A new approach to finding the core for numerical data is discussed. Then a measurement to find whether an attribute is in the core or not is given. This new method for finding the core is used for attribute reduction. It is tested and compared by using eight different machine-learning algorithms. Also, it is discussed how this material is used to rank the importance of attributes in data classification. Finally, the algorithms and codes to convert data to pertinent data and to find the core is also provided.
\end{abstract}


\tableofcontents

\section{Introduction}

Pawlak`s rough set theory \cite{Pawlak3} handles the approximation of sets in terms of equivalence (indiscernibility) relations. The primary application of this theory is data analysis. In Pawlak`s work \cite{Pawlak}, the indiscernibility relations arise when one considers a given set of attributes. Two objects are equivalent if their values of all attributes in the data are the same. Thivagar et al.\cite{Tha} introduce rough topology by means of rough sets and apply it to analyze real-life problems. They find the key attributes of some diseases to decide whether a patient has a disease or not. Several applications of the theory deal with the Boolean type of data in which attributes usually take values yes and no; or 0 and 1. A contribution of this work is to give a generalization of the rough topology and the core to numerical data by reducing data to a usable form by using the standard deviation of attributes. Also, this method provides a new model for attribute reduction for large-scale data processing.

\section{Preliminaries}

Let $U$ be a non-empty set of objects called the universe. A relation $R$ on $U$ is a subset of the cartesian product $U\times U$. An element $(a,b)\in R$ is generally written as $aRb$. A relation on $U$ is called reflexive if $aRa$ for all $a\in U$. It is called symmetric if $aRb$ implies $bRa$ for all $a, b\in U$. It is called transitive if $aRb$ and $bRc$ then $aRc$ for all $a,b,c\in U$. If $R$ is reflexive, symmetric, and transitive, then $R$ is said to be an equivalence relation on $U$.

Let $R$ be an equivalence relation on a set $U$, and let $x\in U$. The set of all elements in $U$ that is related to $x$ is called the equivalence class of $x$ under $R$ and is denoted by $[x]_R$. That is, $[x]_R=\{y\in U| xRy\}$. The set of all equivalence classes $U/R$ of $R$ in $U$ gives a partition of $U$, which means that all equivalence classes are disjoint, and the union of them is $U$.

\begin{definition}\cite{Pawlak} Let $U$ be a non-empty finite set and $R$ be an equivalence relation on $U$. An approximation space is a pair $(U, R)$. Let $X$ be a subset of $U$.
\begin{enumerate}
\item[(i)] The lower approximation of $X$ with respect to $R$ is
$$R_{\star}(X)=\cup_{x\in U}\{x|[x]_R\subset X\}.$$
\item[(ii)] The upper approximation of $X$ with respect to $R$ is 
$$R^{\star}(X)=\cup_{x\in U}\{x|[x]_R\cap X\neq \emptyset\}.$$
\item[(iii)] The boundary region of $X$ with respect to $R$ is
$$B_{R}(X)=R^{\star}(X)-R_{\star}(X).$$
\end{enumerate}
The set $X$ is called a rough set with respect to $R$ if $B_{R}(X)\neq \emptyset$.
\end{definition}

\section{Rough Topology}

In this section, we first introduce the definition of a topology and a basis for a topology. We secondly present the rough topology which is given by Thivagar et al. in \cite{Tha} in terms of the lower and the upper approximations.
\begin{definition} A topology on a set $U$ is a collection $\tau$ of subsets of $U$ satisfying the following properties:
\begin{enumerate}
\item[(T1)] $\emptyset,U\in \tau$.
\item[(T2)] The union of the elements of $\tau$ is in $\tau$.
\item[(T3)] The intersection of the finite number of elements of $\tau$ is in $\tau$.
\end{enumerate}
The pair $(U, \tau)$ is called a topological space.
\end{definition}

Let $(U,\tau)$ be a topological space. A basis for $(U,\tau)$ is a collection $\beta\subset\tau$ such that for each $A\in \tau$ and each $x\in A$, there exists $B\in \beta$ such that $x\in B\subset A$.

\begin{definition}\cite{Tha} Let $U$ be a non-empty finite set and $R$ be an equivalence relation on $U$. For $X\subset U$, $\tau_R=\{U,\emptyset, R_{\star}(X), R^{\star}(X), B_{R}(X) \}$ forms a topology on $U$, which is called a rough topology on $U$ with respect to $X$.
\end{definition}

\begin{lemma} \cite{Tha} The set $\beta_{R}=\{U, R_{\star}(X), B_{R}(X)\}$ is a basis for the rough topology $\tau_R$ on $U$ with respect to $X$.
\end{lemma}

\begin{example}Let $U =\{1,2,3,4,5\}$ and $$R=\{(1,1), (2,2), (3,3), (4,4), (5,5), (1,2), (2,1), (3,5), (5,3) \}$$be an equivalence relation on $U$. Then the set of equivalence classes of $U$ by the equivalence relation $R$ is $U/R=\{\{1,2\},\{3,5\},\{4\}\}$. For $X = \{1,2,3\}$, $R^{\star}(X) =\{1,2,3,5\}, R_{\star}(X) =\{1, 2\}$ and $B_{R}(X) = \{3, 5\}$. Therefore, the rough topology $\tau_R = \{U, \emptyset, \{1,2\}, \{1,2,3,5\},\{3,5\}\}$. The basis for $\tau_R$ is $\beta_R = \{U, \{1,2\}, \{3,5\}\}$.
\end{example}

\begin{definition} \cite{Tha} A subset $M$ of the set of attributes is called the core of $R$ if $\beta_M\neq \beta_{(R-\{r\})\}}$ for every $r$ in $M$. That is, the elements of the core cannot be removed without affecting the classification power of attributes.  
\end{definition}

\begin{example}
Consider the Table \ref{table1}, which is taken from Pawlak \cite{Pawlak2}.
\begin{center}
\begin{longtable}{|p{1cm}||p{2cm}|p{2.8cm}|p{2.3cm}||p{0.6cm}|} 
\caption{Sample Data 1\label{table1}}\\
\hline
Patient&Headache(H)&Muscular pain(M)&Temperature(T)&Flu\\
\hline
\endfirsthead
\multicolumn{5}{c}%
{\tablename\ \thetable\ -- \textit{Continued from previous page}} \\
\hline
Patient&Headache(H)&Muscular pain(M)&Temperature(T)&Flue \\
\hline
\endhead
\hline \multicolumn{5}{r}{\textit{Continued on next page}}\\
\endfoot
\hline
\endlastfoot
  1 & No  & Yes & High & Yes\\
  \hline
  2  & Yes & No & High & Yes\\
  \hline
  3  & Yes  & Yes & Very High & Yes\\
  \hline
  4  & No  & Yes & Normal & No\\
  \hline
  5  & Yes  & No & High & No\\
  \hline
  6  & No  & Yes & Very High & Yes\\
\end{longtable}
\end{center}
Let $U=\{1,2,3,4,5,6\}$ be the set of patients, and $X=\{1,2,3,6\}$ be the set of patients having flue. Let $R$ be an equivalence (indiscernibility) relation, in which two patients are equivalent if their values of all attributes are the same. Then the set of equivalence classes of $U$ by the equivalence relation $R$ is $U/R=\{\{1\},\{2,5\},\{3\}, \{4\}, \{6\}\}$. For $X=\{1,2,3,6\}$, the upper approximation $R^{\star}(X)=\{1,2,3,5, 6\}$, the lower approximation $R_{\star}(X)=\{1, 3, 6\}$, and $B_{R}(X)=\{2, 5\}$. Therefore the rough topology $\tau_R=\{U, \emptyset, \{1,3,6\}, \{1,2,3,5,6\},\{2,5\}\}$. The basis for $\tau_R$ is $\beta_R=\{U, \{1,3,6\}, \{2,5\}\}$.

If we remove the attribute "Headache" from the set of condition attributes, the family of equivalence classes with the resulting set of attributes is given by $U/(R-\{H\})=\{\{1\},\{2,5\},\{3,6\}, \{4\}\}$. Then, the lower and upper approximations of $X$ with respect to $R-\{H\}$ are given by $(R-\{H\})^{\star}(X)$ =\{1,2,3,5,6\} and $(R-\{H\})_{\star}(X)$=\{1,3,6\}, respectively. Therefore, $\tau_{R-\{H\}}$=\{U, $\emptyset$, \{1,3,6\}, \{1,2,3,5,6\}, \{2,5\}\}. The basis for this topology $\tau_{(R-\{H\})}$ is given by $\beta_{(R-\{H\})}=\{U, \{1,3,6\}, \{2,5\}\}$. Therefore, $\beta_R=\beta_{(R-\{H\})}$, which means "Headache" is not in the $Core(X)$.

If the attribute "Muscular pain" is omitted, then $U/(R-\{M\})$=\{ \{1\}, \{2,5\}, \{3\}, \{4\}, \{6\}\}, which is the same with $U/R$. Hence, $\tau_{(R-\{M\})}=\tau_R$ and $\beta_{(R-\{H\})}= \beta_R$, which means "Muscular pain" is not in the $Core(X)$.

If we remove the attribute "Temperature" from the set of condition attributes, the family of equivalence classes with the resulting set of attributes is given by $U/(R-\{T\})=\{\{1,4,6\},\{2,5\},\{3\} \}$. Then, the lower and upper approximations of $X$ with respect to $R-\{T\}$ are given by $(R-\{T\})^{\star}(X)=\{1,2,3,4,5,6\}$ and $(R-\{T\})_{\star}(X)=\{3\}$. Therefore, $\tau_{R-\{T\}}=\{U, \emptyset, \{3\}\}$. The basis for this topology $\tau_{(R-\{T\})}$ is given by $\beta_{(R-\{T\})}=\{U, \{3\}\}$. Therefore, $\beta_R\neq \beta_{(R-\{H\})}$, which means "Temperature" is in the $Core(X)$.

Therefore, $Core(X)=\{\text{Temperature}\}$. If we take $X=\{4,5\}$ as the set of patients not having flu, then similarly $Core(X) = \{\text{Temperature}\}$.\\
\textbf{Observation:} We conclude that "Temperature" is the key attribute to decide whether a patient has flu or not.
\end{example}

\section{The Rough Topology for numerical data}

The rough topology and the core can be used to analyze many real-life problems like diseases, electrical transmission lines, decision-making problems, etc. in the literature \cite{Sayed,Atik,Nasef}. However, the values of attributes are like "yes or no"; or "high, normal, very high". In this section, we give a generalization of the method to analyze numerical data by converting them to pertinent data utilizing standard deviations of the attributes. We also provide algorithms and Python codes to find equivalence classes, the lower and the upper approximations, topologies, bases, and the core. 

Let $(U, R)$ be an approximation space with indiscernibility (equivalence) relation $R$. Two objects are equivalent if and only if their values of all attributes are the same. However, most of the values are different from each other in the numerical data. One needs a method for converting numerical data into formats that help you analyze by using the rough topology and the core. We use standard deviations of attributes to do this. We assume that two objects take the same value if they are close to each other as near as the standard deviation of the attribute. The algorithm of the procedure is as follows:

\begin{algorithm}[h!]\label{Alg1}\caption{}
\begin{enumerate}
\item[Step 1:] Given a data table, columns of which are attributes, rows of which are objects, and entries of the table are attribute values, pick one of the attributes.
\item[Step 2:] Take the maximum ($Max$) and the standard deviation ($St$) of the chosen attribute. 
\item[Step 3:] Assign $1$ to the values between $(Max)$ and $(Max-St)$ and discard these rows. Find the next maximum after discarding rows assigned as $1$. Assign $2$ to the values between the new $(Max)$ and new $(Max-St)$ and discard them. Repeat this process until every value is assigned to a new value.
\item[Step 4:] Repeat Step 2 and Step 3 for every attribute.
\item[Step 5:] Generate the new data table. 
\end{enumerate}
(see the acknowledgment for the link for Python codes).
\end{algorithm}

\begin{example}
Consider the Table \ref{table2}, which is taken from Järvinen \cite{Jar}.
\begin{center}
\begin{longtable}{|p{1cm}||p{2.3cm}|p{3.1cm}|p{2.5cm}||p{1cm}|} 
\caption{Sample Data 2\label{table2}}\\
\hline
Patient&Temperature(T)&Blood Pressure(BP)&Hemoglobin(HB)&Results\\
\hline
\endfirsthead
\multicolumn{5}{c}%
{\tablename\ \thetable\ -- \textit{Continued from previous page}} \\
\hline
Patient&Temperature(T)&Blood Pressure(BP)&Hemoglobin(HB)&Results\\
\hline
\endhead
\hline \multicolumn{5}{r}{\textit{Continued on next page}}\\
\endfoot
\hline
\endlastfoot
  1&39.3&103/65&125&No\\
  \hline
  2&39.1&97/60&116&No\\
  \hline
  3&39.2&109/71&132&No\\
  \hline
  4&37.1&150/96&139&Yes\\
  \hline
  5&37.3&145/93&130&Yes\\
  \hline
  6&37.8&143/95&121&Yes\\
  \hline
  7&36.7&138/83&130&No\\
  \hline
\end{longtable}
\end{center}
By applying the algorithm \ref{Alg1} to Table \ref{table2}, we get the following:

\begin{center}
\begin{longtable}{|p{1cm}||p{2.6cm}|p{3.2cm}|p{2.7cm}||p{1cm}|} 
\caption{Converted Sample Data 2\label{table3}}\\
\hline
Patient&Temperature(T)&Blood Pressure(BP)&Hemoglobin(HB)&Results\\
\hline
\endfirsthead
\multicolumn{5}{c}%
{\tablename\ \thetable\ -- \textit{Continued from previous page}} \\
\hline
Patient&Temperature(T)&Blood Pressure(BP)&Hemoglobin(HB)&Results\\
\hline
\endhead
\hline \multicolumn{5}{r}{\textit{Continued on next page}}\\
\endfoot
\hline
\endlastfoot
1 & 1 & 2 & 2 & No\\
\hline
2 & 1 & 1 & 3 & No\\
\hline
3 & 1 & 2 & 1 & No\\
\hline 
4 & 2 & 2 & 1 & Yes\\
\hline
5 & 2 & 2 & 2 & Yes\\
\hline
6 & 2 & 3 & 3 & Yes\\
\hline
7 & 3 & 1 & 2 & No\\
\end{longtable}
\end{center}
Let $U=\{1,2,3,4,5,6,7\}$ be the set of patients, and $X=\{4,5,6\}$ be the set of patients having a positive result. Let $R$ be an equivalence (indiscernibility) relation, in which two patients are equivalent if their values of all attributes are the same. Then the set of equivalence classes of $U$ by the equivalence relation $R$ is $U/R=\{\{1\}, \{2\}, \{3\}, \{4\}, \{5\}, \{6\}, \{7\} \}$. For $X=\{4, 5, 6\}$, the upper approximation $R^{\star}(X) =\{4,5,6\}$, the lower approximation $R_{\star}(X) =\{4, 5, 6\}$, and $B_{R}(X) =\emptyset$. Therefore the rough topology $\tau_R = \{U, \emptyset, \{4,5,6\} \}$. The basis for $\tau_R$ is $\beta_R = \{U, \{4,5,6\} \}$.

If we remove the attribute "Temperature" from the set of condition attributes, the family of equivalence classes with the resulting set of attributes is given by $U/(R-\{T\})$=\{\{1,5\},\{2\},\{3,4\},\{6\},\{7\} \}. Then, the lower and the upper approximations, and boundary of $X$ with respect to $R-\{T\}$ are given by $(R-\{T\})^{\star}(X) = \{1,3,4,5,6\}$ and $(R-\{T\})_{\star}(X)= \{6\}$, and $B_{(R-\{T\})}(X) = \{1,3,4,5 \}$, respectively. Therefore, $\tau_{R-\{T\}} = \{U, \emptyset, \{1,3,4,5,6\}, \{6\}, \{1,3,4,5 \} \}$. The basis for this topology $\tau_{(R-\{T\})}$ is given by $\beta_{(R-\{T\})} = \{U, \{1,3,4,5 \}, \{6\}\}$. Therefore, $\beta_R\neq \beta_{(R-\{T\})}$, which means "Temperature" is in the $Core(X)$.

If the attribute "Blood Pressure" is omitted, then $U/(R-\{BP\})$=\{ \{1\}, \{2\}, \{3\}, \{4\},\{5\}, \{6\}, \{7\} \}, which is the same with $U/R$. Hence, $\tau_{(R-\{BP\})}=\tau_R$ and $\beta_{(R-\{BP\})}= \beta_R$, which means "Blood Pressure" is not in the $Core(R)$.

If we remove the attribute "Hemoglobin" from the set of condition attributes, the family of equivalence classes with the resulting set of attributes is given by $U/(R-\{HB\})$=\{\{1,3\},\{2\}, \{4,5\}, \{6\},\{7\} \}. Then, the lower and upper approximations of $X$ with respect to $R-\{HB\}$ are given by $(R-\{HB\})^{\star}(X) = \{4,5,6\}$ and $(R-\{HB\})_{\star}(X)= \{4,5,6\}$,respectively. Hence, $\tau_{(R-\{HB\})}=\tau_R$ and $\beta_{(R-\{HB\})}= \beta_R$, which means "Hemoglobin" is not in the $Core(R)$.

Therefore, $Core(X)=\{ \text{Temperature} \}$. If  we take $X=\{1,2,3,7\}$ as the set of patients having a negative result, then similarly $Core(X) = \{\text{Temperature}\}$.\\
\textbf{Observation:} We conclude that "Temperature" is the key attribute to decide whether a patient has a positive result or not.
\end{example}

\section{A topological method for big numerical data}

For big numerical data, the rough topology is restrictive since the topology could be changed by only one object (row). In other words, even if only one object (row) is in the boundary and the lower approximation is different from the boundary or lower approximation for the equivalence relation which is omitted one attribute, then topologies are different. As a result, most attributes are at the core of big data. However, we can give a measurement for this change (see definition \ref{measure}). For example, let`s consider the data with 1000 objects. If only 10 objects change the boundary and the lower approximation, these differences are possibly insignificant, so it could be tolerated for big data. This tool also provides the order of importance of attributes in the dataset.

\begin{definition}
Let $(U, R)$ be an approximation space and $M$ be the set of attributes. For $r\in M$ and $X\subset U$, the accuracy of an attribute and the accuracy of the core with respect to $X$ are defined respectively by 
$$\mu_r(X)=\frac{|B_R(X)|}{|B_{(R-r)}(X)|}$$
$$\mu_{RT}(X)=Sup_{r\in M}\{\frac{|B_R(X)|}{|B_{(R-r)}(X)|}\}.$$
\end{definition}

\begin{definition}\label{measure} Let $(U,R)$ be an approximation space and $M$ be the set of attributes. For $r\in M$ and $X\subset U$, the accuracy of the boundary for $r$ with respect to $X$ are defined by
$$\nu_r(X)=| B_R(X) - B_{(R-r)}(X) |$$.
\end{definition}

\begin{remark}$\mu_{RT}(X)=\mu_{RT}(U-X)$ and $\nu_{r}(X)=\nu_{r}(U-X)$ for any $r\in M$.
\end{remark}

Obviously, $\mu_{RT}(X)\leq 1$. $\nu_{r}(X)$ is a measuring tool used to lay out insignificant objects that can be tolerated. For example, if $\nu_r(X)\leq 10$, then one can consider that an attribute $r$ is not in the core because it could be tolerated for the data with $1000$ objects. It is a method for determining which attributes are more important than others in this classification. It also lists the relative importance of these attributes.
\subsection{Applications}

\begin{example}\label{cyro} The dataset was obtained from the UCI Machine Learning Repository. This dataset contains information about wart treatment results of 90 patients using cryotherapy \cite{cryo}.
Seven attributes are sex, age, time, number of warts, type, area, and result of treatment.
\begin{center}
\begin{longtable}{|p{1.1cm}||p{0.6cm}|p{0.6cm}|p{0.7cm}|p{2.7cm}|p{0.7cm}|p{0.6cm}|p{1cm}|} 
\caption{Cryotherapy Treatment\label{table4}}\\
\hline
Objects&Sex&Age&Time&Number of Warts&Type&Area&Results\\
\hline
\endfirsthead
\multicolumn{5}{c}%
{\tablename\ \thetable\ -- \textit{Continued from previous page}} \\
\hline
Objects&Sex&Age&Time&Number of Warts&Type&Area&Results\\
\hline
\endhead
\hline \multicolumn{5}{r}{\textit{Continued on next page}}\\
\endfoot
\hline
\endlastfoot 
  1 & 1	& 35 & 12 & 5 & 1 & 100 & 0\\
  \hline
  2 & 1 & 29 & 7 & 5 & 1 & 96 & 1\\
  \hline
  3 & 1 & 50 & 8 & 1 & 3 & 132 & 0\\
  \hline
  4 & 1 & 32 & 11.75 & 7 & 3 & 750 & 0\\
  \hline
  5 & 1 & 67 & 9.25 & 1 & 1	& 42 & 0\\
  \hline
  6 & 1	& 41 & 8 & 2 & 2 & 20 & 1\\
  \hline
  7 & 1 & 36 & 11 & 2 & 1 & 8 & 0\\
  \hline
  8 & 1 & 59 & 3.5 & 3 & 3 & 20 & 0\\
  \hline
  9 & 1 & 20 & 4.5 & 12 & 1 & 6 & 1\\
  \hline
  10 & 2 & 34 & 11.25 & 3 & 3 & 150 & 0\\
  \hline
  \vdots  & \vdots  & \vdots & \vdots & \vdots & \vdots &\vdots &  \vdots \\
  \hline
\end{longtable}
\end{center}
By applying the algorithm \ref{Alg1} to Table \ref{table4}, we get the following:
\begin{center}
\begin{longtable}{|p{1.1cm}||p{0.6cm}|p{0.6cm}|p{0.7cm}|p{2.7cm}|p{0.7cm}|p{0.6cm}|p{1cm}|} 
\caption{Converted Cryotherapy Treatment\label{table5}}\\
\hline
Objects&Sex&Age&Time&Number of Warts&Type&Area&Results\\
\hline
\endfirsthead
\multicolumn{5}{c}%
{\tablename\ \thetable\ -- \textit{Continued from previous page}} \\
\hline
Objects&Sex&Age&Time&Number of Warts&Type&Area&Results\\
\hline
\endhead
\hline \multicolumn{5}{r}{\textit{Continued on next page}}\\
\endfoot
\hline
\endlastfoot 
  1 & 2	& 3 & 1 & 2 & 3 & 2 & 0\\
  \hline
  2 & 2 & 3 & 2 & 2 & 2 & 2 & 1\\
  \hline
  3 & 2 & 2 & 2 & 3 & 1 & 2 & 0\\
  \hline
  4 & 2 & 3 & 1 & 2 & 1 & 1 & 0\\
  \hline
  5 & 2 & 1 & 1 & 3 & 3	& 2 & 0\\
  \hline
  6 & 2	& 2 & 2 & 3 & 2 & 3 & 1\\
  \hline
  7 & 2 & 3 & 1 & 3 & 3 & 3 & 0\\
  \hline
  8 & 2 & 1 & 3 & 3 & 1 & 3 & 0\\
  \hline
  9 & 2 & 4 & 3 & 1 & 3 & 3 & 1\\
  \hline
  10 & 1 & 3 & 1 & 3 & 1 & 2 & 0\\
  \hline
  \vdots  & \vdots  & \vdots & \vdots & \vdots & \vdots &\vdots &  \vdots \\
  \hline
\end{longtable}
\end{center}
Let $U=\{1, 2, 3,\cdots, 88, 89, 90\}$ be the set of objects, and $X$ and $Y$ be the set of objects having result $0$ and $1$, respectively. Let $R$ be an equivalence (indiscernibility) relation, in which two objects are equivalent if their values of all attributes are the same. If attributes which have $\nu_r(X)\leq 2$ are eliminated, then the $Core(X)=\{\text{time},\text{number \ of \ warts}, \text{area} \}$ (See Table \ref{Ex1}). Similarly, $Core(U-X)=Core(Y)=\{\text{time},\text{number \ of \ warts}, \text{area} \}$.
\begin{center}
\begin{longtable}{|p{1.6cm}||p{0.3cm}|p{0.4cm}|p{0.5cm}|p{2cm}|p{0.5cm}|p{0.5cm}|} 
\caption{The Core of the Dataset \label{Ex1}}\\
\hline
Column&1&2&3&4&5&6\\
\hline
\endfirsthead
\multicolumn{7}{c}%
{\tablename\ \thetable\ -- \textit{Continued from previous page}} \\
\hline
Column&1&2&3&4&5&6\\
\hline
\endhead
\hline \multicolumn{7}{r}{\textit{Continued on next page}}\\
\endfoot
\hline
\endlastfoot
Attribute(r)&Sex&Age&Time&N. of Warts&Type&Area\\
\hline
$\nu_r(X)$&1&2&21&5&0&6\\
\end{longtable}
\end{center}
As a result, key attributes to decide the result of the treatment are time, number of warts, and area. The order of significance of these attributes for the classification is time, area, and number of warts.

In this part, we apply 8 different machine learning algorithms, which are Support Vector Classifier (SVC), Random Forest Classifier (RFC), Linear Regression (LR), Gradient Boosting Classifier (GBC), Extreme Gradient Boosting (XGBC), Linear Discriminant Analysis (LDA), Gaussian Naive Bayes (GNB), and Hybrid (HYB), to the data by using all attributes and attributes in the core, respectively. Here, the Hybrid algorithm is the algorithm that is created based on whether 4 out of 7 algorithms predict correctly or not. Then, the results are compared for each method and each class. Looking at average classification accuracy, we get better results by using the core.

Machine learning algorithms are used by default setting.  It is carried out without any optimizations or parameters using a random selection procedure. We split the data set so that 80\% is used to train the model and 20\% is used to test with a fixed random selection. We consider running the ML algorithms multiple times, comparing the average result, and then reporting metrics.

The bar chart Figure \ref{Bar1} illustrates the accuracy rates of several machine learning algorithms based on two sets of features: "Core Attributes" and "All Attributes." Across the majority of algorithms, using the "Core Attributes" yields a higher accuracy compared to "All Attributes," with notable examples including the SVC (0.556 vs 0.500) and HYB (0.889 vs 0.722) models. However, in the case of the RFC algorithm, the difference in accuracy between the two feature sets is marginal (0.833 vs 0.778), suggesting that adding more attributes did not significantly improve or harm its performance.
For other algorithms such as LR, GBC, XGBC, and LDA, the "Core Attributes" consistently outperform "All Attributes" with a difference of about 0.1 in some instances, suggesting the core features contribute more meaningfully to the predictive power of these models. Overall, the "Core Attributes" feature set appears to perform better across most models, which may indicate a more focused and optimized feature selection. However, certain models like LDA and GNB show lower overall performance compared to others.

\begin{figure}[htbp]
    \centering
    \begin{tikzpicture}
        \begin{axis}[
            ybar,
            bar width=20pt,
            width=\textwidth,
            height=0.5\textwidth,
            enlarge x limits=0.1,
            legend style={at={(0.5,-0.2)}, anchor=north, legend columns=-1},
            ylabel={Accuracy Rate},
            symbolic x coords={SVC, RFC, LR, GBC, XGBC, LDA, GNB, HYB},
            xtick=data,
            ymin=0.45, ymax=0.92,
            nodes near coords,
            every node near coord/.append style={
                font=\small, 
                /pgf/number format/.cd, 
                fixed,
                fixed zerofill,
                precision=3
            }, % Ensure full precision (3 decimal places) at the top of the bars
            xlabel={ML Algorithms}
        ]
        % Core Attributes data (blue)
        \addplot[style={fill=blue}] coordinates {(SVC,0.5) (RFC,0.833) (LR,0.889) (GBC,0.722) (XGBC,0.889) (LDA,0.889) (GNB,0.778) (HYB,0.889)};
        % All Attributes data (orange)
        \addplot[style={fill=orange}] coordinates {(SVC,0.556) (RFC,0.778) (LR,0.778) (GBC,0.778) (XGBC,0.778) (LDA,0.667) (GNB,0.668) (HYB,0.722)};
       
        \legend{Core Attributes, All Attributes}
        \end{axis}
    \end{tikzpicture}
    \caption{Comparison of Core Attributes and All Attributes for ML Algorithms \label{Bar1}}
\end{figure}

The following Table \ref{MLA1} demonstrates the classification outcomes of the eight ML algorithms.
\newpage
\begin{center}
\begin{longtable}{|p{1.3cm}||p{3cm}|p{3cm}|} 
\caption{ML Algorithm Results \label{MLA1}}\\
\hline
Method& Use all attributes & Use the core\\
\hline
\endfirsthead
\multicolumn{3}{c}%
{\tablename\ \thetable\ -- \textit{Continued from previous page}} \\
\hline
Method & Use all attributes & Use the core\\
\hline
\endhead
\hline \multicolumn{3}{r}{\textit{Continued on next page}}\\
\endfoot
\hline
\endlastfoot
SVC & Accuracy:0.556 & Accuracy:0.5 \\
 & Precision:0.529 & Precision:0.5 \\  
 & Recall:1 & Recall:1 \\ 
 & F1:0.692 & F1:0.667  \\
\hline
RFC & Accuracy:0.778 & Accuracy:0.889 \\
 & Precision:0.727 & Precision:1 \\  
 & Recall:0.889 & Recall:0.778\\ 
 & F1:0.8 & F1:0.875 \\
\hline
LR & Accuracy:0.778 & Accuracy:0.889\\
 & Precision:0.857 & Precision:1\\  
 & Recall:0.667 & Recall:0.778\\ 
 & F1:0.75 & F1:0.875\\
\hline
GBC & Accuracy:0.778 & Accuracy:0.833 \\
 & Precision:0.727 & Precision:0.8 \\  
 & Recall:0.889 & Recall:0.889 \\ 
 & F1:0.8 & F1:0.842 \\
\hline
XGBC & Accuracy:0.778 & Accuracy:0.889 \\
 & Precision:0.727 & Precision:1 \\  
 & Recall:0.889 & Recall:0.778 \\ 
 & F1:0.8 & F1:0.875 \\
\hline
GNB & Accuracy:0.667 & Accuracy:0.778 \\
 & Precision:0.636 & Precision:0.778 \\  
 & Recall:0.778 & Recall:0.778 \\ 
 & F1:0.7 & F1:0.778 \\
\hline
LDA & Accuracy:0.667 & Accuracy:0.889 \\
 & Precision:0.667 & Precision:1 \\  
 & Recall:0.667 & Recall:0.778 \\ 
 & F1:0.667 & F1:0.875 \\
\hline
HYB & Accuracy:0.722 & Accuracy:0.889 \\
 & Precision:0.667 & Precision:1 \\  
 & Recall:0.889 & Recall:0.778 \\ 
 & F1:0.762 & F1:0.875 \\
\hline
\end{longtable}
\end{center}

The class-wise distribution of a classification model's predicted performance is called a confusion matrix. The confusion matrices we generated for the hybrid algorithm (which achieves the highest accuracy with Core attributes) employing all of the features and the features in the core are shown below in Figures \ref{Con2} and \ref{Con1}, respectively.

\begin{figure}[htbp]
\centering
\begin{tikzpicture}
    \begin{axis}[
            colormap={bluewhite}{color=(white) rgb255=(90,96,191)},
            xlabel=Predicted,
            xlabel style={yshift=0pt},
            ylabel=Actual,
            ylabel style={yshift=0pt},
            xticklabels={0,1}, % changed
            xtick={0,1}, % changed
            xtick style={draw=none},
            yticklabels={0, 1}, % changed
            ytick={0, 1}, % changed
            ytick style={draw=none},
            enlargelimits=false,
            colorbar,
            xticklabel style={
              rotate=90
            },
            nodes near coords={\pgfmathprintnumber\pgfplotspointmeta},
            nodes near coords style={
                yshift=-7pt
            },
        ]
        \addplot[
            matrix plot,
            mesh/cols=2, % changed
            point meta=explicit,draw=gray
        ] table [meta=C] {
            x y C
            0  0 9
            1  0 0
            
            0 1 2
            1 1 7
        }; % added every entry where x=4 or y=4
    \end{axis}
\end{tikzpicture}
\caption{Confusion Matrix for the Hybrid Algorithm (The Core) \label{Con1}}
\end{figure}

\begin{figure}[htbp]
\centering
\begin{tikzpicture}
    \begin{axis}[
            colormap={bluewhite}{color=(white) rgb255=(90,96,191)},
            xlabel=Predicted,
            xlabel style={yshift=0pt},
            ylabel=Actual,
            ylabel style={yshift=0pt},
            xticklabels={0,1}, % changed
            xtick={0,1}, % changed
            xtick style={draw=none},
            yticklabels={0, 1}, % changed
            ytick={0, 1}, % changed
            ytick style={draw=none},
            enlargelimits=false,
            colorbar,
            xticklabel style={
              rotate=90
            },
            nodes near coords={\pgfmathprintnumber\pgfplotspointmeta},
            nodes near coords style={
                yshift=-7pt
            },
        ]
        \addplot[
            matrix plot,
            mesh/cols=2, % changed
            point meta=explicit,draw=gray
        ] table [meta=C] {
            x y C
            0  0 5
            1  0 4
            
            0 1 1
            1 1 8
        }; % added every entry where x=4 or y=4
    \end{axis}
\end{tikzpicture}
\caption{Confusion Matrix for the Hybrid Algorithm (All Attributes) \label{Con2}}
\end{figure}

\end{example}

\subsection{An Application to Dataset with Categorical Decision Attribute} It can also be applied to data with that a decision attribute takes more than two values. One can get better classification results by using the core rather than using all attributes by applying machine learning algorithms to most numerical data. Also, with $\nu_r(X)$, in the classification, the most important attributes and their importance ranking in deciding which class an object is in can be given.

Let $U$ be the set of objects, and $X_1, X_2,\cdots, X_n$ be the set of objects having result $1, 2,\cdots, n$, respectively. Let $R$ be an equivalence (indiscernibility) relation, in which two objects are equivalent if their values of all attributes are the same. Then, we have $Core(X_i)=Core(U-X_i)$ for all $i=1, 2,\cdots, n$. However, $Core(U-X_i)$ does not need to equal to $Core(U-X_j)$ for all $1\leq i,j\leq n$. Having said that, the most important attributes and their relative weights in determining the class an object belongs to can be determined using $\nu_r(X_i)$ in the classification process for all $i=1, 2,\cdots, n$.

\begin{example}\label{Seeds} The dataset was obtained from the UCI Machine Learning Repository about measurements of geometrical properties of three different types of wheat kernels: Kama, Rosa, and Canadian, $70$ objects each \cite{seeds}. In the data, seven attributes of wheat kernels were measured:
\begin{enumerate}
\item[]A: area,
\item[]B: perimeter,
\item[]C: compactness,
\item[]D: length of kernel,
\item[]E: width of kernel,
\item[]F: asymmetry coefficient,
\item[]G: length of kernel groove,
\item[]Result: kernel type (Kama=1, Rosa=2, Canadian=3).
\end{enumerate}
\begin{center}
\begin{longtable}{|p{1.1cm}||p{0.8cm}|p{0.8cm}|p{1cm}|p{0.8cm}|p{0.8cm}|p{0.8cm}|p{0.8cm}||p{1cm}|} 
\caption{Wheat Kernel Data\label{table7}}\\
\hline
Objects&A&B&C&D&E&F&G&Results\\
\hline
\endfirsthead
\multicolumn{5}{c}%
{\tablename\ \thetable\ -- \textit{Continued from previous page}} \\
\hline
Objects&A&B&C&D&E&F&G&Results\\
\hline
\endhead
\hline \multicolumn{5}{r}{\textit{Continued on next page}}\\
\endfoot
\hline
\endlastfoot 
  1 & 15.26 & 14.84 & 0.871 & 5.763 & 3.312 & 2.221 & 5.22 & 1
\\
  \hline
  2 & 14.88 & 14.57 & 0.8811 & 5.554 & 3.333 & 1.018 & 4.956 & 1\\
  \hline
  3 & 14.29 & 14.09 & 0.905 & 5.291 & 3.337 & 2.699 & 4.825 & 1\\
  \hline
  4 & 13.84 & 13.94 & 0.8955 & 5.324 & 3.379 & 2.259 & 4.805 & 1\\
  \hline
  5 & 16.14 & 14.99 & 0.9034 & 5.658 & 3.562 & 1.355 & 5.175 & 1\\
  \hline
  \vdots  & \vdots  & \vdots & \vdots & \vdots & \vdots &\vdots &  \vdots & \vdots \\
  \hline
  71 & 17.63 & 15.98 & 0.8673 & 6.191 & 3.561 & 4.076 & 6.06 & 2\\
  \hline
  72 & 16.84 & 15.67 & 0.8623 & 5.998 & 3.484 & 4.675 & 5.877 & 2\\
  \hline
  73 & 17.26 & 15.73 & 0.8763 & 5.978 & 3.594 & 4.539 & 5.791 & 2\\
  \hline
  74 & 19.11 & 16.26 & 0.9081 & 6.154 & 3.93 & 2.936 & 6.079 & 2\\
  \hline
  75 & 16.82 & 15.51 & 0.8786 & 6.017 & 3.486 & 4.004 & 5.841 & 2\\
  \hline
  \vdots  & \vdots  & \vdots & \vdots & \vdots & \vdots &\vdots &  \vdots & \vdots \\
  \hline
  141 & 13.07 & 13.92 & 0.848 & 5.472 & 2.994 & 5.304 & 5.395 & 3\\
  \hline
  142 & 13.32 & 13.94 & 0.8613 & 5.541 & 3.073 & 7.035 & 5.44 & 3\\
  \hline
  143 & 13.34 &	13.95 & 0.862 &	5.389 &	3.074 &	5.995 &	5.307 & 3\\
  \hline
  144 &	12.22 &	13.32 &	0.8652 & 5.224 & 2.967 & 5.469 & 5.221 & 3\\
  \hline
  145 &	11.82 &	13.4  & 0.8274 & 5.314 & 2.777 & 4.471 & 5.178 & 3\\
  \hline
  \vdots  & \vdots  & \vdots & \vdots & \vdots & \vdots &\vdots &  \vdots & \vdots \\
  \hline
\end{longtable}
\end{center}
By applying the algorithm \ref{Alg1} to the Table \ref{table7}, we get the following:

\begin{center}
\begin{longtable}{|p{1.1cm}||p{0.8cm}|p{0.8cm}|p{1cm}|p{0.8cm}|p{0.8cm}|p{0.8cm}|p{0.8cm}||p{1cm}|} 
\caption{Converted Wheat Kernel Data\label{table8}}\\
\hline
Objects&A&B&C&D&E&F&G&Results\\
\hline
\endfirsthead
\multicolumn{5}{c}%
{\tablename\ \thetable\ -- \textit{Continued from previous page}} \\
\hline
Objects&A&B&C&D&E&F&G&Results\\
\hline
\endhead
\hline \multicolumn{5}{r}{\textit{Continued on next page}}\\
\endfoot
\hline
\endlastfoot 
  1 & 2 & 2 & 2 & 3 & 2 & 4 & 3 & 1
\\
  \hline
  2 & 3 & 3 & 2 & 3 & 2 & 5 & 4 & 1\\
  \hline
  3 & 3 & 3 & 1 & 4 & 2 & 4 & 4 & 1\\
  \hline
  4 & 3 & 3 & 1 & 4 & 2 & 4 & 4 & 1\\
  \hline
  5 & 2 & 2 & 1 & 3 & 2 & 5 & 3 & 1\\
  \hline
  \vdots  & \vdots  & \vdots & \vdots & \vdots & \vdots &\vdots &  \vdots & \vdots\\
  \hline
  71 & 2 & 1 & 3 & 2 & 2 & 3 & 1 & 2\\
  \hline
  72 & 2 & 2 & 3 & 2 & 2 & 3 & 2 & 2\\
  \hline
  73 & 2 & 2 & 2 & 2 & 2 & 3 & 2 & 2\\
  \hline
  74 & 1 & 1 & 1 & 2 & 1 & 4 & 1 & 2\\
  \hline
  75 & 2 & 2 & 2 & 2 & 2 & 3 & 2 & 2\\
  \hline
  \vdots  & \vdots  & \vdots & \vdots & \vdots & \vdots &\vdots &  \vdots & \vdots\\
  \hline
  141 & 3 & 3 & 3 & 3 & 3 & 2 & 3 & 3\\
  \hline
  142 & 3 & 3 & 3 & 3 & 3 & 1 & 3 & 3\\
  \hline
  143 & 3 &	3 & 3 &	3 &	3 &	2 &	3 & 3\\
  \hline
  144 &	3 &	3 &	3 & 4 & 3 & 2 & 3 & 3\\
  \hline
  145 &	4 &	3  & 4 & 4 & 4 & 3 & 3 & 3\\
  \hline
  \vdots  & \vdots  & \vdots & \vdots & \vdots & \vdots &\vdots &  \vdots & \vdots\\
  \hline
\end{longtable}
\end{center}
Let $U=\{1, 2, 3,\cdots, 208, 209, 210\}$ be the set of objects, and $X$=\{1, 2, 3, $\cdots$, 68, 69, 70\}, $Y$=\{71, 72 , 73, $\cdots$, 138, 139, 140\} and $Z$=\{141, 142, 143, $\cdots$, 208, 209, 210\} be the set of objects having result $1$, $2$, and $3$, respectively. Let $R$ be an equivalence (indiscernibility) relation, in which two objects are equivalent if their values of all attributes are the same.

\begin{table}[h]
\centering
\caption{The Core of Dataset}
\begin{tabular}{|c|c c c c c c c|}
\hline
\textbf{Column} & 1 & 2 & 3 & 4 & 5 & 6 & 7 \\
\hline
\textbf{Attribute (r)} & A & B & C & D & E & F & G \\
\hline
$\nu_r(X)$ & 0 & 1 & 25 & 1 & 11 & 67 & 55 \\
\hline
$\nu_r(Y)$ & 0 & 0 & 4 & 1 & 0 & 23 & 52 \\
\hline
$\nu_r(Z)$ & 0 & 1 & 21 & 0 & 11 & 44 & 3 \\
\hline
\end{tabular}
\end{table}

As a consequence, firstly, $Core(X)=\{C, E, F, G\}$. The order of importance of the attributes for the objects classified as in $X$ is $F, G, C, E$. Secondly, $Core(Y)=\{C, F, G\}$. The order of importance of the attributes for the objects classified as in $Y$ is $G, F, C$. Lastly, $Core(Z)=\{C, E, F, G\}$. The order of importance of the attributes for the objects classified as in $Z$ is $F, C, E, G$. Finally, key attributes to decide the type of seeds are $C, E, F $, and $G$ columns.
\end{example}

\section{Conclusion and Future Directions}

In this work, we give the rough topology and core for numerical data. By using the core, one can also get better results by using fewer attributes (attributes in the core) for machine learning algorithms. As a result, the rough topology is a useful model for attribute reductions for numerical data. This method could be applied to big numerical data for future selection problems in future research. We will use this method to make predictions of qualities such as maintenance cost overruns, work accidents, and the severity of construction quality failures for datasets.

Another project is to apply the rough topology to solve the missing values problem in incomplete numerical information tables. The rough topological method for numerical data could be given like Salama's work for Boolean Type of data \cite{salama, Yang}.

Finally, this new rough topological method can be given for expansions of rough sets in incomplete information systems by taking tolerance relations rather than indiscernibility relations to apply missing value problems for numerical data \cite{Krys, Yang}.

\acknowledgements \label{Ack}
I would like to thank Kenan Evren Boyabatlı for helping me to write Python codes.\\
\textbf{Code Availability Statement:}
Some or all models, or codes that support the findings of this study are available from the corresponding author upon reasonable request. See the link URL:
\begin{center}
\url{https://github.com/EvReN-jr/TDR-Topological-Dimensional-Reduction}\label{link}
\end{center}
for Python and C++ codes to find the converted data, the approximations, topology, and the Core.

\bibliographystyle{fundam}
\bibliography{math}

\end{document}